\documentclass[aps]{revtex4}
\usepackage{graphicx}
\begin{document}

\title{Universality of Regge and vibrational trajectories in a
semiclassical model}
\author{B. Silvestre-Brac
}
\affiliation{Institut des Sciences Nucl\'{e}aires, 
Avenue des Martyrs 53, F-38026 Grenoble-Cedex, FRANCE}
\author{F. Brau
\thanks{Chercheur I.I.S.N., 
E-mail: fabian.brau@umh.ac.be} 
and C. Semay
\thanks{Chercheur qualifi\'{e} F.N.R.S., 
E-mail: claude.semay@umh.ac.be}
}
\affiliation{Groupe de Physique Nucl\'eaire Th\'eorique, Universit\'e
de Mons-Hainaut, B-7000 Mons, Belgium}
\date{\today}

\begin{abstract}
The orbital and radial excitations of light-light mesons
are studied in the framework of the dominantly orbital state
description. The equation of motion is characterized by a
relativistic kinematics supplemented by the usual funnel potential
with a mixed scalar and vector confinement. The influence of finite
quark masses and potential parameters on Regge and vibrational
trajectories is discussed. The case of heavy-light mesons is also
presented.
\end{abstract}

\maketitle

\section{Introduction}
\label{sec:intro}

\begin{figure}
\label{fig:regge}
\centering
\includegraphics*[width=10cm]{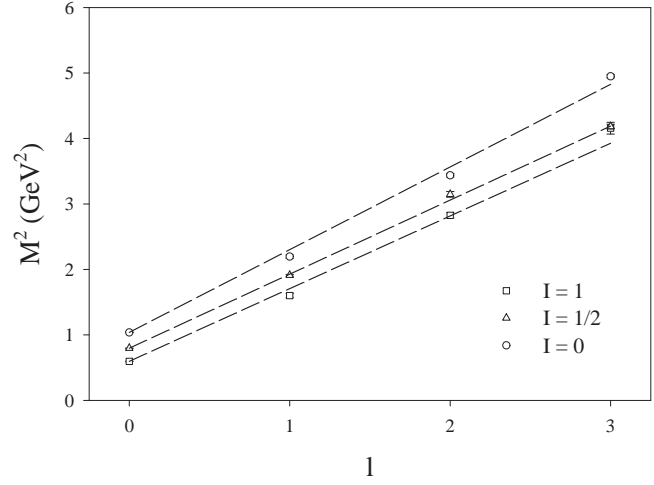}
\protect\caption{Square mass $M^2$ of some light-light mesons as a
function of their assumed total orbital angular momentum $l$. The
isovector $n \bar n$ mesons ($\rho$ family) are indicated by square
boxes, strange mesons ($K^\star$ family) by triangles and $s \bar s$
mesons ($\phi$ family) by circles. The
mesons represented are characterized by an internal total spin quantum
number equal to one. Excepted for the $l=0$ states, each mass is given
by the center of gravity of mesons with the same $l$ but with total spin
different (see Sec.~\protect\ref{sec:exp}). The remarkable linear
dependence is pointed out by straight
dashed lines.}
\end{figure}

One of the most striking feature of the light-light meson spectra is
certainly the existence of the so-called ``Regge trajectories'', that
is to say the linear dependence of the square mass as a function of the
total spin quantum number. One can see in Fig.~\ref{fig:regge} the
remarkable linear trajectories of the $\rho$, $K^\star$ and $\phi$
families. It is well known
that this behavior can be obtained in potential models by using a
confinement potential well adjusted for the kinematics. In
non-relativistic descriptions of meson spectra, a $r^{2/3}$ confining
potential must be used \cite{fabr88}, while in relativistic
kinematics the linear Regge trajectories are obtained with a linear
confinement \cite{luch91} and results also naturally from the
relativistic flux tube model of mesons \cite{laco89,sema95}. It is worth
noting that this mass behavior is also expected for heavy-light mesons.

\par An interesting theoretical approach of light-light and heavy-light
mesons is given by the dominantly orbital state (DOS) description
\cite{goeb90} for which the leading Regge trajectory is a ``classical"
result while radially excited states can be treated semiclassically. In
a recent paper \cite{olss97}, this DOS model has been applied to study
light-light and heavy-light meson spectra with a confinement potential
being a mixture of scalar and vector components. In the case of one or
two quarks with vanishing masses, it was shown that linear Regge
(orbital) and ``vibrational'' (radial) trajectories are obtained for an
arbitrary scalar-vector mixture but that the ratio of radial to orbital
energies is strongly dependent on the mixture.

\par In this paper, we extend the result of Ref.~\cite{olss97} for
light-light and heavy-light mesons by considering the presence of quarks
with finite masses and the introduction of a Coulomb-like interaction
and constant potential in addition to the confinement. In
the limit of small masses and small strength of the short range
potential, or in the limit of large angular momentum, all calculations
can be worked out analytically. One could expect that the small quark
masses and Coulomb-like potential do not need to be taken into account
to study high orbitally excited state. But we show here that such
effects are interesting to point out. The square mass formula for
light-light mesons composed of two identical quarks as a function of
quantum numbers and potential parameters is established and discussed in
Sec.~\ref{sec:model}. A comparison of our theoretical predictions with
data is performed in Sec.~\ref{sec:exp}. Some concluding remarks are
given in Sec.~\ref{sec:conc}.

\section{The model}
\label{sec:model}

\subsection{basic ideas}
\label{ssec:idea}

The starting point of our model is essentially the same as the one
presented by Olsson \cite{olss97}. Consequently, we just remind the
basic ideas and skip a lot of technical details which can be found in
that reference. In all our formulae, we use the natural units
$\hbar=c=1$. 

\par We consider a system composed of two spinless
particles with the same mass $m$, which are allowed to rotate around
their center of mass fixed at the origin, and to vibrate in the
radial dimension. The Lagrange coordinates are the interdistance $r$
and the rotation angle $\varphi$, angle between an arbitrary fixed
direction and the line joining the two particles. The free
relativistic Lagrangian is very easy to write and depends on the
Lagrange coordinates and their time derivatives.
In order to handle the dynamics between the particles, we
introduce phenomenologically an interaction potential which depends
on the radial coordinate $r$ only. In fact, there exist two Lorentz
structures for this potential: a scalar term $S(r)$ which simply
modifies the mass ($m \rightarrow m+S(r)/2$), and a vector term $V(r)$
which is simply subtracted
from the Lagrangian. Consequently the starting Lagrangian is
given by 
\begin{equation}
\label{2.1}
L(r,\varphi,\dot{r},\dot{\varphi}) =
-\left(2m+S(r)\right)\sqrt{1-\frac{1}{4}\dot{r}^2-\frac{1}{4}
\dot{\varphi}^2r^2} - V(r).
\end{equation}
This formula can be applied to particles with spin in a very crude
manner by considering spin dependent potential terms. 
One can see that $L$ is independent of $\varphi$. This means that the
angular momentum $J=\frac{\partial L}{\partial \dot{\varphi}}$ is a good
quantum number. Taking into account this condition, the Hamiltonian
of the system can be written as 
\begin{equation}
\label{2.2}
H(r,p_r,J)=\sqrt{4p_r^2+\frac{4J^2}{r^2}+\left(2m+S(r)\right)^2}+V(r).
\end{equation}
The idea of the model is to make a classical approximation by
considering uniquely the classical circular orbits (lowest energy
states with given $J$), defined by $r$=$r_0$, and thus $\dot{r}=0$
and $p_r$=0. Let us denote 
\begin{equation}
\label{2.3}
M_0(J)=H(r=r_0,p_r=0,J).
\end{equation}
By using the condition $\frac{\partial L}{\partial r}(r=r_0)=0$ and
performing a number of analytical transformations (see
Ref.~\cite{olss97}),
one finds the fundamental equations (with the conventions $S_0 =
S(r_0)$, and so on) 
\begin{equation}
\label{2.4}
M_0(J)=\frac{r_0}{2} \left[V^\prime_0+2\frac{V_0}{r_0} +
\sqrt{{V^\prime_0}^2+4\frac{(2m+S_0)}{r_0^2}\left[(2m+S_0)+S^\prime_0
r_0\right]}
\right]
\end{equation}
and
\begin{equation}
\label{2.5}
J=\frac{r_0^2}{2} \left[\frac{1}{2}V^\prime_0\left(V^\prime_0+
\sqrt{{V^\prime_0}^2+4\frac{(2m+S_0)}{r_0^2}\left[(2m+S_0)+S^\prime_0
r_0\right]}
\right)+\frac{(2m+S_0)S^\prime_0}{r_0} \right]^{1/2}.
\end{equation}
Equation~(\ref{2.5}) (which is in general a transcendental equation)
allows to calculate $r_0(J)$. Once this value is obtained, it is put
into Eq.~(\ref{2.4}) to get $M_0(J)$.
Up to now, we are in position to obtain only the ground state energy
for a given $J$. In order to get the radial excitations, it is useful
to make a harmonic approximation around the previous classical
orbits. Here again, the details can be found in Ref.~\cite{olss97}. The
harmonic quantum energy is given by
\begin{equation}
\label{2.6}
\Omega(J)=2 \frac{\left[ \left(M_0(J)-V_0\right)V_0^{\prime \prime}
+{S^\prime_0}^2+(2m+S_0)S_0^{\prime
\prime}+\displaystyle{\frac{12J^2}{r_0^4}}-
{V^\prime_0}^2 \right]^{1/2}}{M_0(J)-V_0}.
\end{equation}
Then the mass of the system with an orbital excitation $J$ and a
radial excitation $n$ (0, 1, \ldots) is given, within this
approximation, by 
\begin{equation}
\label{2.7}
M(J,n)=M_0(J)+\Omega(J)\left(n+\frac{1}{2}\right).
\end{equation}
In this general formulation, one sees that there exists, in principle,
a coupling between the orbital and the radial excitations.
In relativistic descriptions of two particle systems (for example in
the meson sector), it is generally the square of the mass which is
considered as the function of the orbital and radial quantum
numbers. If one assumes that $\Omega(J) \ll M_0(J)$ (see
Ref.~\cite{olss97}), then one can write 
\begin{equation}
\label{2.7p}
M^2(J,n) = M_0^2(J) + M_0(J)\Omega(J)(2n+1).
\end{equation}

\subsection{The meson case}
\label{ssec:mes}

\subsubsection{The potential}
\label{sssec:pot}

In the previous subsection, we gave a formalism applicable to any type
of potential. Such a study could be undertaken from the numerical point
of view, by solving the equations giving $r_0$ and
reporting the value in the expression of $M_0$ and $\Omega$. However
there are some drawbacks to proceed that way. First, because of non
linearity, there can exist several solutions for Eq.~(\ref{2.5}) and it
may be difficult to discriminate between them. Second, the physical
analysis is less obvious when the results originate from a numerical
treatment. So, it is highly desirable to study a model for which an
analytical solution is available. 

\par In Ref.~\cite{olss97}, the author studied the spectra of the mesons
or, more precisely, the Regge trajectories for a system composed of a
quark and an antiquark with $m=0$ interacting via a linear confinement
potential that is partly scalar and partly vector. He showed that a
linear behavior $M_0^2(J) \propto J$ automatically emerges. 

\par In this paper, we revisit this study, but in a more general
framework. First, we relax the constraint of zero mass and consider the
mass of the quarks $m$ as a free parameter. Second, we admit a long
range linear confinement supplemented by a short range Coulomb part and
constant potential. Indeed quantum chromodynamics tells us that it
should be so. The one-gluon exchange process gives rise to a Coulomb
term of vector type, while multigluon exchanges are responsible for the
linear confinement. Since we ignore the Lorentz structure of the
confinement, we suppose, as in Ref.~\cite{olss97}, that it is partly
scalar and partly vector. The importance of each one is reflected
through a parameter $f$ whose value is 0 for a pure vector, and 1 for a
pure scalar. 

\par Thus, in this paper, we apply the previous formalism, with
potentials given by 
\begin{equation}
\label{2.8}
S(r)=f\:a\:r
\end{equation}
in which $a$ is the usual string tension, whose value should be
around 0.2 GeV$^2$, and 
\begin{equation}
\label{2.9}
V(r)=(1-f)\:a\:r\:-\:\frac{\kappa}{r}
\end{equation}
in which $\kappa$ is proportional to the strong coupling constant
$\alpha_s$. A reasonable value of $\kappa$ should be in the range 0.1 to
0.6. We will show below that we are able to obtain an analytical
solution for this rather realistic situation.

\subsubsection{Scaling properties}
\label{sssec:scp}

In our case, we are faced with a number of parameters, whose origin
is very different
\begin{itemize}
\item the mass $m$ (in GeV), the string constant $a$ (in
$\rm{GeV}^2$) and the Coulomb constant $\kappa$ (dimensionless) are
dynamical parameters;
\item the $f$ parameter (dimensionless) reflects our ignorance about the
Lorentz structure of the confinement;
\item the quantum numbers $J$ and $n$ of the state.
\end{itemize}
In what follows, we will be interested by the variation of
$M^2(m,a,\kappa,f,J,n)$ as a function of quantum numbers $J$
and $n$, and for a given set of physical parameters $m,a,\kappa,f$.
$M_0$ does not
depend on $n$, as well as $\Omega$, so that
the $n$ dependence is rather obvious (see Eq.~(\ref{2.7p})).
Let us focus on the dependence on $J$ and on the physical parameters.
Actually, it is possible to exploit a scaling law in our equation. One
can remark that
$\sqrt{a}$ is an energy unit, and $\frac{1}{\sqrt{a}}$ a length unit,
so that it is advantageous to work with these natural units. Let us
define reduced dimensionless quantities, labeled with an overhead
bar, relative to these units. Explicitly, we have
\begin{equation}
\label{2.10}
\overline{m}=\frac{m}{\sqrt{a}}  \: ; \:
\overline{M}_0=\frac{M_0}{\sqrt{a}}  \: ; \:
\overline{M}=\frac{M}{\sqrt{a}}  \: ; \:
\overline{\Omega}=\frac{\Omega}{\sqrt{a}}  \: ; \:
\overline{r}_0=r_0 \sqrt{a}. 
\end{equation}
If we put these values into our fundamental Eq.~(\ref{2.5}), with
application to our particular potentials (\ref{2.8}-\ref{2.9}), we
find, after some manipulations, the equation for $\overline{r}_0$
\begin{eqnarray}
\label{2.11}
&&f^2(2f-1)\, \overline{r}_0^8+4 \overline{m} f(2f-1)\,
\overline{r}_0^7+2\left[2 \overline{m}^2 (2f-1)-\kappa
f^2 (1-f)\right]\, \overline{r}_0^6-
8\kappa \overline{m} f (1-f)\, \overline{r}_0^5
\nonumber\\ 
&&-\left[4(1-f) \Big((1-f)J^2+2\kappa \overline{m}^2\Big)+f^2 \Big(
\kappa^2 +8J^2\Big)\right]\, \overline{r}_0^4 - 4 \overline{m} f
\left(\kappa^2+4J^2\right)\,
\overline{r}_0^3 \nonumber\\
&&-4\kappa \left[\kappa \overline{m}^2+2J^2 (1-f)\right]\,
\overline{r}_0^2 +4J^2
\left(4J^2-\kappa^2\right)=0.
\end{eqnarray}
Equation~(\ref{2.4}) giving the value of $M_0$ leads in this case to
\begin{equation}
\label{2.12}
\overline{M}_0=\frac{(1-2f)\, \overline{r}_0^4-2\overline{m} f\,
\overline{r}_0^3 +4J^2-\kappa^2}{\overline{r}_0\left((1-f)\,
\overline{r}_0^2+\kappa\right)}.
\end{equation}
Similarly, Eq.~(\ref{2.6}) giving the value of
$\Omega$ leads to
\begin{eqnarray}
\label{2.13}
&&\overline{\Omega}= \nonumber\\
&&2\, \left\{\left[(1-f)(2f-1)\,
\overline{r}_0^6+3\kappa (2f-1)\, \overline{r}_0^4+4\kappa
\overline{m} f\, \overline{r}_0^3+
3(1-f)(4J^2-\kappa^2)\,
\overline{r}_0^2+(4J^2-\kappa^2)\kappa\right]\right.
\nonumber \\
&&\left.\left[(1-f)\, \overline{r}_0^2+\kappa\right]
\right\}^{1/2}/\left\{\overline{r}_0\, \left[-f^2\, \overline{r}_0^4-2
\overline{m} f\, \overline{r}_0^3+4J^2\right]\right\}.
\end{eqnarray}
Some comments are in order. 
\begin{itemize}
\item We see that the use of this scaling law allows to reduce the
number of parameters to four ($\overline{m},\kappa,f,J$) in the
dimensionless quantities. 
\item If both $m$=0 and $\kappa$=0, it remains only the powers (0,4,8)
in Eq.~(\ref{2.11}) so that the equation is always soluble by using
$\overline{r}_0^4$ as a variable; one can see that only one relevant
physical solution exists. It should be stressed that this case
corresponds to the study of (Ref.~\cite{olss97}); we checked that our
general formalism applied to that peculiar case gives the solution
presented in that reference.
\item The complicated equation (\ref{2.5})
simplifies in our particular case to a polynomial of eighth order in
$\overline{r}_0$ (Eq.~(\ref{2.11})). In principle, there could exist 8
solutions, but we must reject non physical complex and real negative
values, and the real positive values which lead to negative values for
$\overline{M}_0$ and $\overline{\Omega}$. Starting from the unique
physical solution
$\overline{r}_0(\overline{m}=0,\kappa=0)$, we can expect, by continuity,
that there still remains only one physical solution for small values of
$\overline{m}$ and $\kappa$. Although we are not able to prove that the
physical solution is always unique, it seems that this is a general
property; we remarked that this was always true for a
wide range of parameters and for the cases for which an analytical
solution can be found (see below). 
\item The linear term in $\overline{r}_0$ is always missing in all
expressions.
\item The special values $f=0$ (pure vector confinement) and
$f=1/2$ (half scalar-half vector) lead for Eq.~(\ref{2.11}) to a
polynomial of 6th order only. Moreover, in the case $f=0$, the
polynomial is of 3th order in $r_0^2$. 
\item Odd powers of $\overline{r}_0$ disappear in Eq.~(\ref{2.11}) when
$\overline{m}=0$. By choosing $\overline{r}_0^2$ as a
variable, we are left with a 4th order polynomial.
\end{itemize}
Although the things look quite nice now, it is possible to simplify
further, using another scaling law.
As one will see, this new scaling law is very powerful and allows to
derive universal conclusions concerning Regge trajectories. Indeed,
it is possible to eliminate the $J$ dependence with help of the
following changes of variables 
\begin{equation}
\label{2.14}
x = \overline{r}_0\:J^{-1/2}  \; ; \;
u = \overline{m}\:J^{-1/2}  \; ; \;
v = \kappa\:J^{-1}.
\end{equation}
Using these variables, our polynomial (\ref{2.11}) is rewritten
\begin{eqnarray}
\label{2.15}
&&f^2(2f-1)\, x^8+4 u f(2f-1)\, x^7+2\left[2 u^2 (2f-1)
-v f^2 (1-f)\right]\, x^6- 8v u f (1-f)\, x^5
\nonumber\\ 
&&-\left[4(3 f^2-2 f+1)+8(1-f)u^2 v+ f^2 v^2\right]\, x^4 - 4 u f
(v^2+4)\,x^3 \nonumber\\
&&-4v \left[v u^2+2(1-f)\right]\,x^2 +4(4-v^2)=0.
\end{eqnarray}
The interest of this equation is that the variable $x$ now depends
only on 3 variables $x=x(f,u,v)$. Alternatively, one can put
Eq.~(\ref{2.12}) under the form 
\begin{eqnarray}
\label{2.16}
\overline{M}_0&=&J^{1/2} g(f,u,v),  \nonumber \\
g(f,u,v)&=&\frac{(1-2f)\, x^4-2u f\,x^3 +4-v^2}
{x\left[(1-f)\,x^2+v\right]},
\end{eqnarray}
and equation (\ref{2.13}) under the form 
\begin{eqnarray}
\label{2.17}
&&\overline{\Omega}=2 J^{-1/2} h(f,u,v),  \nonumber \\
&&h(f,u,v)= \nonumber \\ 
&&\left\{\left[(1-f)(2f-1)\,x^6+3v (2f-1)\,x^4
+4v u f\, x^3+3(1-f)(4-v^2)\, x^2+(4-v^2)v\right]\right. \nonumber \\
&&\left.\left[(1-f)\, x^2+v\right]
\right\}^{1/2} / 
\left\{x\, \left[-f^2\, x^4-2u f\,x^3+4\right]\right\}.
\end{eqnarray}
Under these forms, the formalism is the most general and the simplest
possible. Except very specific cases that we have studied, it is not
possible to get an analytical solution for this general formulation. To
proceed further, we need to make some kind of approximations.

\subsubsection{Regge and vibrational trajectories}
\label{sssec:rgt}

\begin{figure}
\label{fig:coef}
\centering
\includegraphics*[width=10cm]{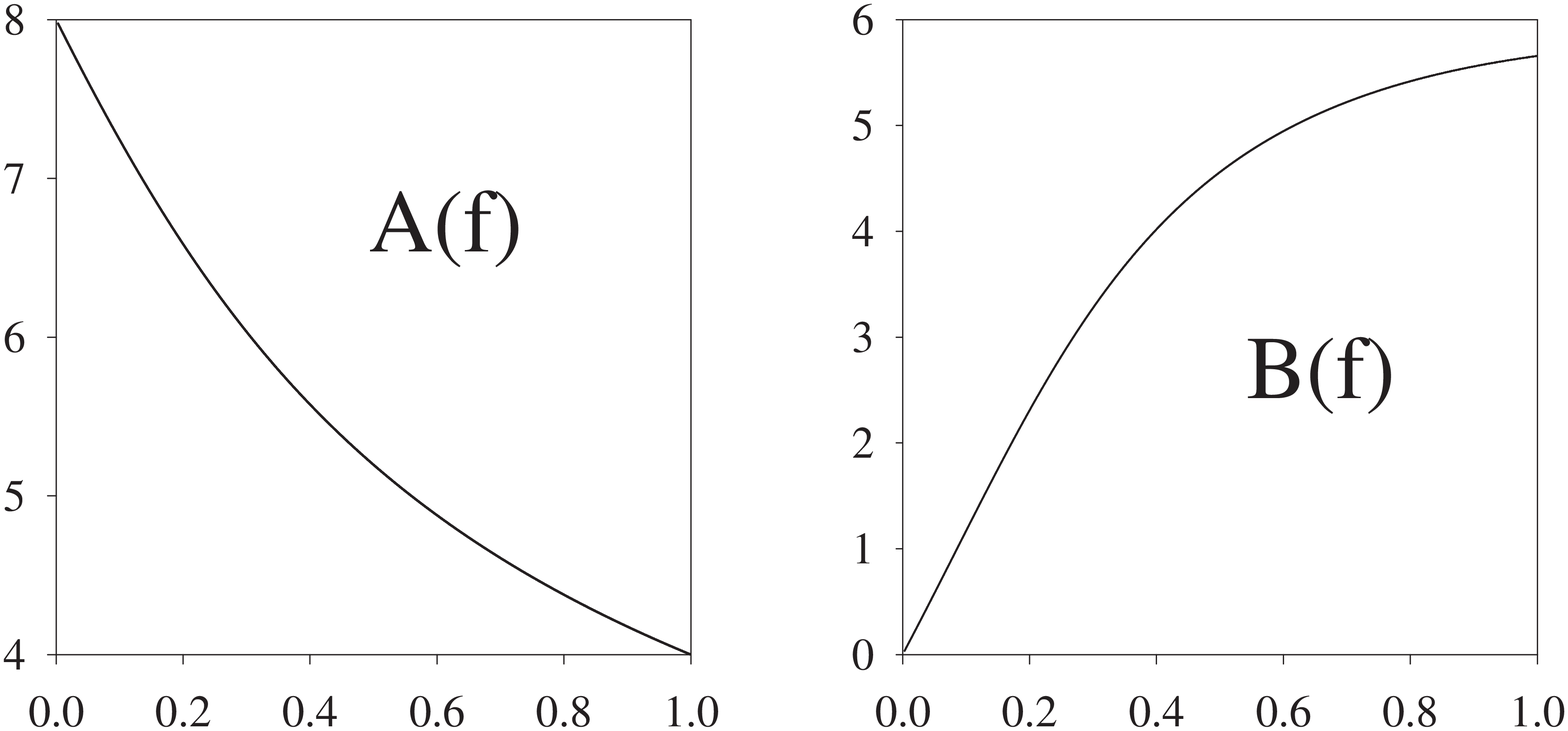}
\includegraphics*[width=10cm]{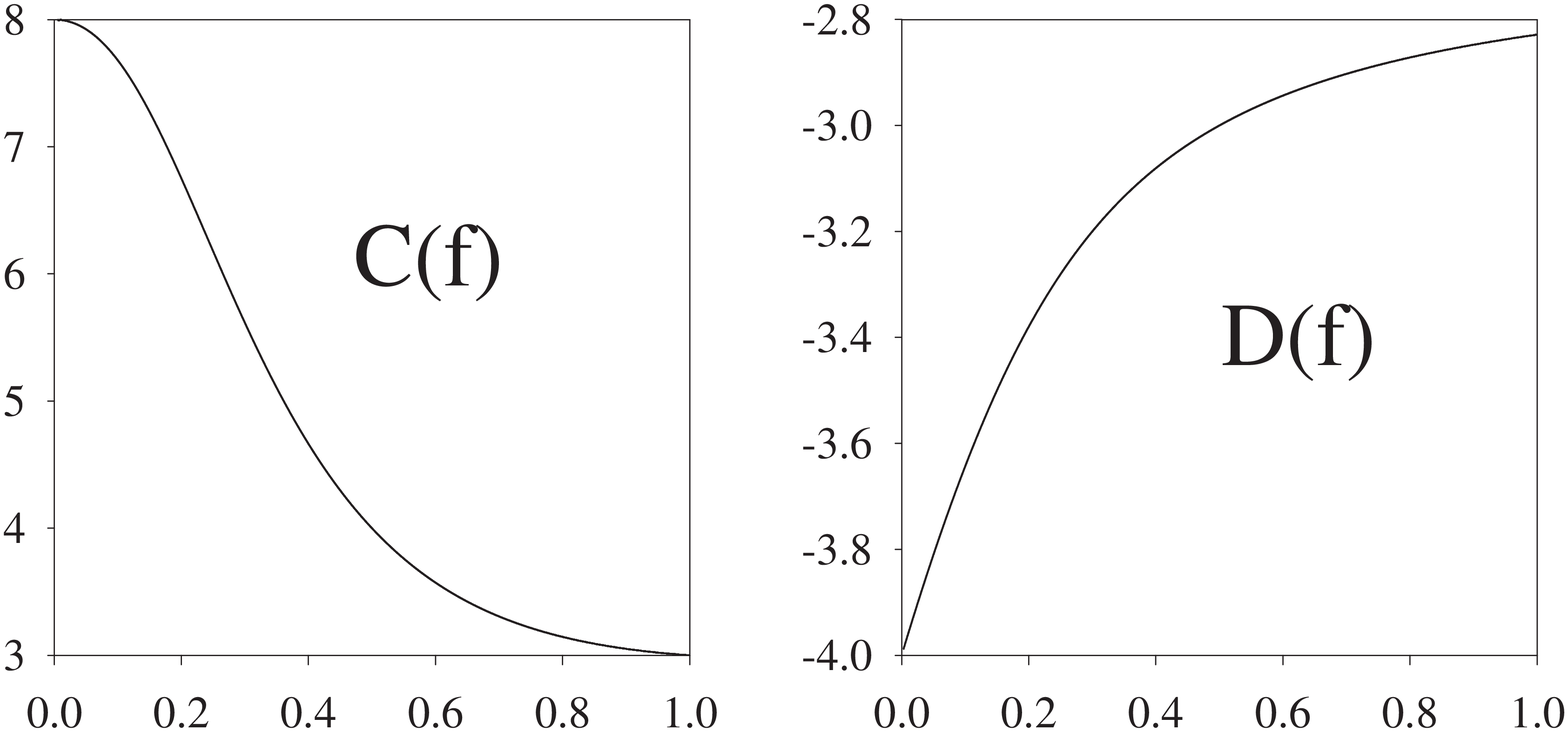}
\includegraphics*[width=10cm]{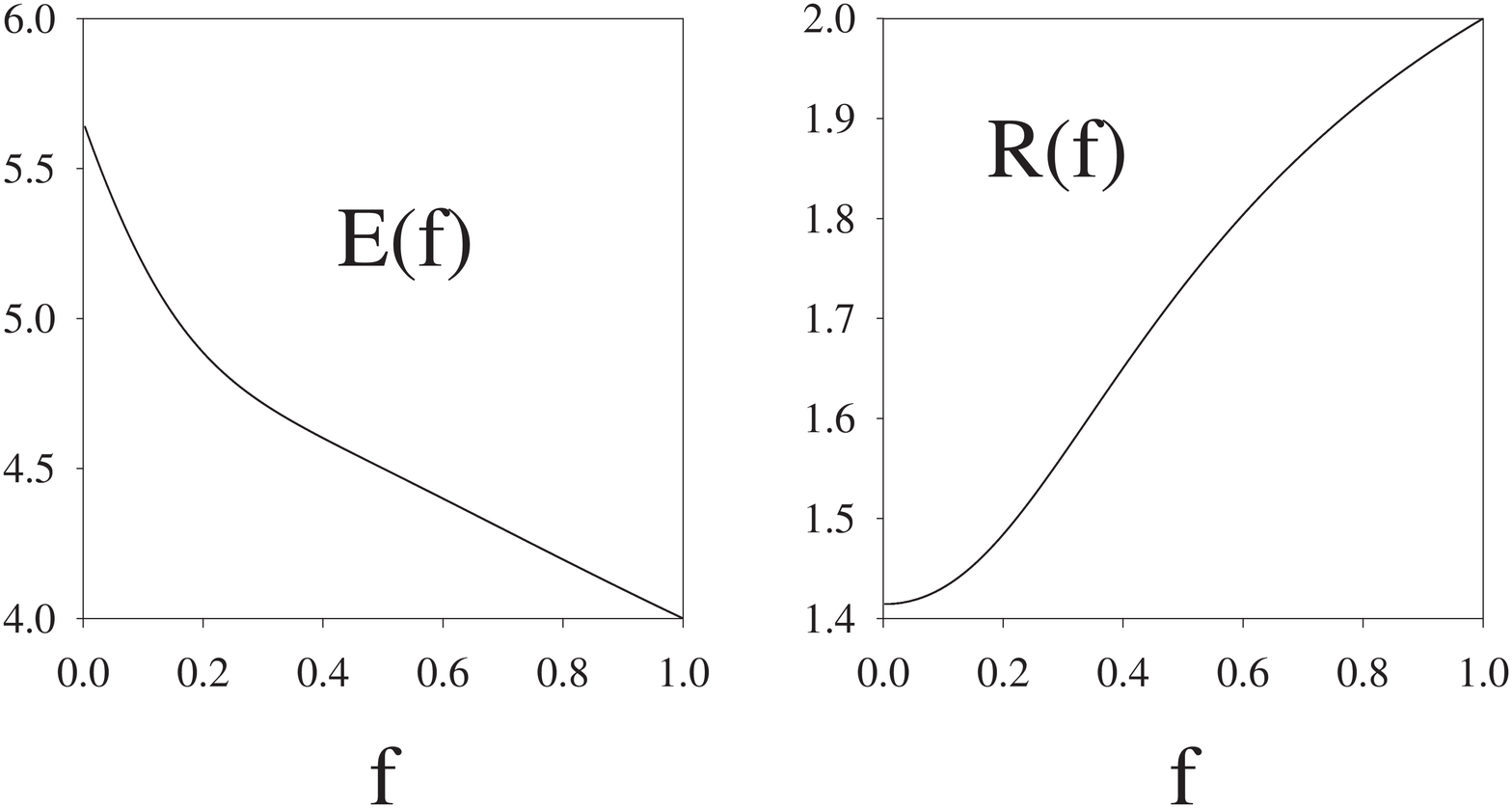}
\protect\caption{Coefficients giving the square mass (\ref{2.19}) as a
function of parameter $f$. The ratio $R(f) = 2E(f)/A(f)$ is also given.}
\end{figure}

From now on, we are interested only in Regge trajectories which give
the behavior of $M^2$ in term of $J$ for large value of $J$. It is
thus natural to consider the values of $u$ and $v$ defined by
Eq.~(\ref{2.14}) as small quantities and to develop $g$ and $h$
functions, given by Eqs.~(\ref{2.16}) and (\ref{2.17}), up
to order $J^{-1}$ that is to say keeping constant terms, terms with
$u$, terms with $u^2$ and terms with $v$. Such a calculation is very
painful and we have been helped a lot by the MATHEMATICA software. We
skip below all the cumbersome details to retain only the important
things.
Basically we expand $x$ as 
\begin{equation}
\label{2.18}
x=\alpha(f)+\beta(f)\,u+\gamma(f)\,u^2+\delta(f)\,v.
\end{equation}
We put this expression into Eq.~(\ref{2.15}) up to the given order and
identify to zero to get the expressions of
$\alpha(f)$, $\beta(f)$, $\gamma(f)$ and $\delta(f)$. Then we report the
resulting
value of $x$ in $g(f,u,v)$ and $h(f,u,v)$, limited again to the desired
order to get $\overline{M}_0$ and $\overline{\Omega}$. Lastly we used
these quantities in Eq.~(\ref{2.7}) to obtain $M$. It is important to
point out that the expression~(\ref{2.7p}) is valid up to our order (the
term in $\Omega^2$ is of higher order). At the end of this study, the
behavior for large $J$ can be traced back, after reintroducing
dimensioned quantities, under the exact form 
\begin{equation}
\label{2.19}
{M}^2=a\,A(f)\, J+B(f)\, m\, \sqrt{a\, J} +C(f)\,m^2+a\, D(f) \kappa 
+a\ E(f) (2n+1) + O(J^{-1/2}).
\end{equation}
The coefficients $A$, $B$, $C$, $D$ and $E$ can be obtained
analytically. Another
coefficient $R(f)=\frac{2\,E(f)}{A(f)}$ is specially important and
will be discussed below. Let us define the auxiliary functions
$s(f),t(f),y(f)$ by 
\begin{eqnarray}
\label{2.20}
s(f)&=&(1-f)^2+2f^2=1-2f+3f^2, \nonumber\\
t(f)&=&\sqrt{s(f)+6f^2}, \nonumber\\
y(f)^4&=&\frac{8}{s(f)+(1-f)\, t(f)}.
\end{eqnarray}
Our coefficients are given by
\begin{eqnarray}
\label{2.21}
A(f)&=&\frac{y^2}{4}\left[t+3(1-f)\right]^2, \nonumber\\
B(f)&=&\frac{y}{f}\left[(1+f)(3f-1)+t\, (1-f)\right], \nonumber\\
C(f)&=&\frac{1}{f^2 t}\left[t\, (s+f^2)+(1-f)(2f-1)\right], \nonumber\\
D(f)&=&-\left[t+3(1-f)\right], \nonumber\\
E(f)&=&A(f) \sqrt{\frac{t}{t+1-f}}, \nonumber \\
R(f)&=&2\sqrt{\frac{t}{t+1-f}}. 
\end{eqnarray}
We plot the various coefficients in Fig.~\ref{fig:coef}. All these
quantities are monotonic functions of $f$ and their ranges from $f=0$ to
$f=1$ are given below
\begin{itemize}
\item $8 \geq A(f) \geq 4$;
\item $0 \leq B(f) \leq 4\sqrt{2}$;
\item $8 \geq C(f) \geq 3$;
\item $-4 \leq D(f) \leq -2\sqrt{2}$;
\item $4\sqrt{2} \geq E(f) \geq 4$;
\item $\sqrt{2} \leq R(f) \leq 2$.
\end{itemize}
Equation~(\ref{2.19}) is the exact expression valid for large $J$; we
succeeded in getting an analytical expression in our model. It is
very powerful since it is universal and the various dependencies are
explicitly seen. One can make the following comments.
\begin{itemize}
\item The dominant term for large $J$ is linear, so that we recover
pure Regge trajectories. As far as the string tension is
consider as a constant, the slope of the Regge trajectories $aA(f)$
is universal in the sense that it is independent of the system
(it is also independent on the strong coupling constant). Only the
confinement drives the behavior; moreover the slope depends on $f$,
the percentage of scalar confinement.
\item The slope of the vibrational trajectories $2aE(f)$ presents the
same characteristics. Even more important, the ratio of the slopes for
radial and orbital trajectories $R(f)$ depends only on $f$. This ratio
is greater than 1, as indicated by experimental data. \item The next to
dominant term behaves like $\sqrt{J}$ and deforms the straight
trajectories for low values of $J$. It must be emphasized that this term
is due only to a non zero mass and is independent of the strong coupling
constant. Moreover, this term is absent whatever the system if the
confinement is purely of vector type, since $B(0)=0$.
\item The displacement of the trajectories from the origin has three
contributions; one $C(f)m^2$ is due to the mass, another
$aD(f)\kappa$ is due to the strong coupling constant and the last one
$aE(f)$ reflects the zero point motion of the harmonic vibration. The
position above or below the origin depends upon the relative
importance of each contribution (do not forget that $C$ and $E$ are
always $>0 $ but that $D$ is always $<$0). The zero point energy of the
orbital motion cannot obviously be calculated in our model. 
\item There is no coupling between orbital and radial motion for
large $J$ values (absence of terms $nJ$). This is only a consequence
of the Coulomb+linear nature of the quark-antiquark potential. This
may not be true for other types of potentials.
\end{itemize}

\subsection{Addition of a constant term}
\label{ssec:const}

It is well known that, in traditional spectroscopy relying upon
Schr\"{o}dinger or spinless-Salpeter equation, it is necessary to add a
constant term to the potential in order to get the absolute values of
the spectra. In those models, the effect of this constant is just to
shift the absolute spectrum keeping the same relative spectrum. One can
raise the question of adding a constant potential in our model. We will
see that it is not so difficult to answer this question in our
framework.

\par In principle it is possible to add a constant $C_{\text{V}}$ to the
vector potential and a constant $C_{\text{S}}$ to the scalar potential.
Let us discuss separately these two cases.

\par If we add a constant $C_{\text{V}}$ to $V(r)$, this does not affect
the functions $V^\prime(r)$, $S(r)$, and $S^\prime(r)$. From
Eq.~(\ref{2.5}), it is clear that the value $r_0$ remains unchanged, and
thus all the quantities depending on $r_0$ remain unchanged except $V_0$
which is change to $V_0+C_{\text{V}}$. Consequently, from
Eq.~(\ref{2.4}), $M_0$ is changed to $M_0+C_{\text{V}}$ and $\Omega$
remains unchanged since it depends only on $M_0-V_0$ (see
Eq.~(\ref{2.6})). Then $M$ is changed to $M+C_{\text{V}}$. Squaring this
quantity and keeping only the terms to good order, the net effect of
$C_{\text{V}}$ is to add supplementary terms to the $B$ and $C$
coefficients.

\par Let us now add a constant term $C_{\text{S}}$ to $S(r)$; $V(r)$ and
all the derivatives are unchanged. Since, in Eq.~(\ref{2.5}), $J$
depends on $S(r)$, $r_0$ does depend on $C_{\text{S}}$. But we see that
it is always the value $\left(2m+S(r)\right)$ which appears. If we
introduce a new mass $m'=m+C_{\text{S}}/2$, we see that the new value of
$r_0$ is simply $r_0(m',J)$ instead of $r_0(m,J)$. Making this
replacement in $M_0$ and $\Omega$, we see that we have now $M_0(m',J)$
and $\Omega(m',J)$ instead of the non-primed values. So, the only
modification in our formalism is just a change from $m$ to $m'$. But in
doing so, we must consider the new value of $u$ defined with the
modified mass $m'$ for applying a limited expansion in term of $J$
($m'/\sqrt{a}$ can be small with respect to 1, while $m$ can be large
with respect to $\sqrt{a}$). Assuming that such an expansion is
justified, the only effect of $C_{\text{S}}$ is to modify the terms in
the $\sqrt{J}$ and the constant contributions.

\par One can now gather both effects into a single formula. Let us
define the dimensionless quantities $\epsilon=1+\frac{C_{\text{S}}}{2m}$
and $\eta=\frac{C_{\text{V}}}{m}$. The only effect of adding the
constants $C_{\text{V}}$ and $C_{\text{S}}$ is to change, in the
universal equation (\ref{2.19}), the $B(f)$ and $C(f)$ coefficients by
new ones $B^\prime(f,\epsilon,\eta)$ and $C^\prime(f,\epsilon,\eta)$
defined by 
\begin{eqnarray}
\label{2.22}
B^\prime(f,\epsilon,\eta)&=&\epsilon \, B(f)+2 \eta \, \sqrt{A(f)},
\nonumber \\
C^\prime(f,\epsilon,\eta)&=& \epsilon^2 \, C(f)+ \eta^2+
2 \epsilon \, \eta \, \frac{t-(1-f)}{2f}.
\end{eqnarray}
In particular, the $A$, $D$ and $E$ coefficients are unchanged, so that
the slopes of the orbital and radial trajectories are still unchanged in
this more elaborated potential, and appears really as universal
quantities depending only on scalar-vector mixture in the confinement.

\subsection{Heavy-light mesons}
\label{ssec:heavy}

The same formalism can be applied to the case of heavy-light mesons in
the limit of infinite heavy quark mass, as mentioned in
Ref.~\cite{olss97}. In light-light mesons considered here, two identical
quarks orbit at the same distance $r/2$ from the center of mass. For a
heavy-light meson, the light quark orbits around the fixed infinitely
massive quark at the distance $r$. To obtain the heavy-light
Hamiltonian, one must perform the following replacements in the
light-light Hamiltonian (\ref{2.2})
\begin{itemize}
\item $\displaystyle{\frac{r}{2}} \rightarrow r$;
\item $2m+S(r) \rightarrow m+S(r)$;
\item $V(r) \rightarrow V(r)$.
\end{itemize}
After calculation, one obtains
\begin{equation}
\label{2.23}
{M}^2=\frac{a}{2}\,A(f)\, J+\frac{1}{2^{3/2}}\,B(f)\, m\, \sqrt{a\, J}
+C(f)\,\frac{m^2}{4}+a\, D(f)
\kappa 
+a\ E(f) (2n+1) + O(J^{-1/2}).
\end{equation}

\section{Discussion of the model}
\label{sec:exp}

\begin{figure}
\label{fig:m2comp}
\centering
\includegraphics*[width=10cm]{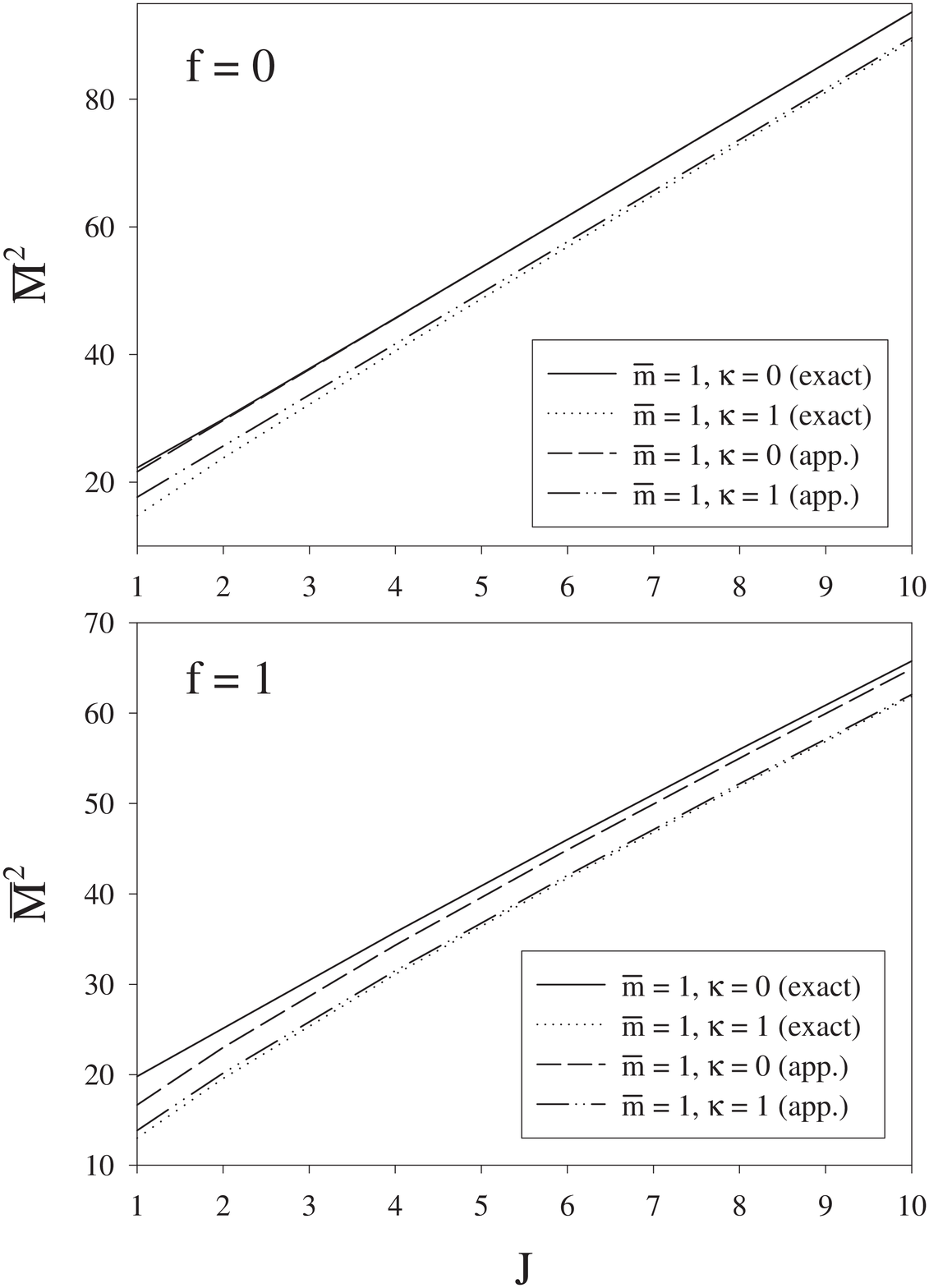}
\protect\caption{Comparison of the reduced square mass $\overline{M}^2$
given by formulae~(\ref{2.7p},\ref{2.11}-\ref{2.13}) (exact), and the
approximate one
given by Eq.~(\ref{2.19}) (app.), as a function of the total angular
momentum
$J$ for values of $f$, $\overline{m}$ and $\kappa$ equal to 0 and 1, and
$C_{\text{S}} = C_{\text{V}} = 0$.} 
\end{figure}

Formula~(\ref{2.19}) gives the meson square mass dependence as a
function of parameters. One can ask whether it is possible to extract
from this relation values for the parameters by comparing with available
data. The value generally considered for the string tension $a$ is
around $0.2 \pm 0.03$ GeV$^2$. Taking $m_u \approx m_d \approx 0.200$
GeV and $m_s \approx 0.450$ GeV, we can see that the parameter
$\overline{m}$ runs from 0.5 to 1. The Coulomb strength $\kappa$ is not
exactly known, but in potential models a range from 0.1 to 0.6 is
currently accepted. So we have to check that formula~(\ref{2.19}) can be
applied for this range of parameters. In Fig.~\ref{fig:m2comp}, we
compare the reduced square mass $\overline{M}^2$ calculated exactly by
formulae~(\ref{2.7p},\ref{2.11}-\ref{2.13}) with the one given by
Eq.~(\ref{2.19}), as a function of the total angular momentum $J$ for
some values of $f$, $\overline{m}$, $\kappa$ and with $C_{\text{S}} =
C_{\text{V}} = 0$. One can expect that the quality of the approximation
is better for small $\overline{m}$ and $\kappa$ parameters. This is
precisely what we get. We present the figure for $\overline{m}=1$, a
value already large and, nevertheless, the approximation still works
reasonably well. For $f=0$, the approximate value is a pure straight
line ($B(0)=0$). The deviation is maximum for large $\kappa$ values, as
expected. For $f=1$, there is a small deformation both for the exact
solution and the approximate one. Moreover, the curvature is the same in
both cases. Note that the concavity of the curve may be inverted by
introducing non-vanishing values for $C_{\text{S}}$ and/or
$C_{\text{V}}$. For large values of $J$, it is apparent that our
approximate expression is always good. For small values of $J$, the
deviation is maximum for $\kappa=0$, but we are in a region
($\overline{m}$ large and $J$ small) for which we expect some error. By
continuity, other values of $f$ give intermediate situations.

\par We see that, even in the rather unfavorable case, the approximate
expression does not differ so much from the exact results. Consequently,
we will base our further discussion on Eq.~(\ref{2.19}). This last
equation can rewritten in the form
\begin{equation}
\label{3.1}
M^2(J,n) = {\cal A}\,J + {\cal B}\,\sqrt{J} + {\cal C} + {\cal D}\,n
\end{equation}
with ${\cal A}$, ${\cal B}$, ${\cal C}$ and ${\cal D}$ depending on
parameters $f$, $a$, $m$, $\kappa$, $C_{\text{S}}$ and $C_{\text{V}}$.
Knowing three orbital excitations, for instance $M^2(J,n)$, $M^2(J+1,n)$
and $M^2(J+2,n)$, it is possible to calculate ${\cal A}$ and ${\cal B}$
quantities. Adding a radial excitation, for instance $M^2(J,n+1)$, the
quantities ${\cal C}$ and ${\cal D}$ can also be calculated. The
formulae are given by
\begin{eqnarray}
\label{3.2}
{\cal A} &=& K_{1,0}(J)\,M^2(J+2,n) - K_{2,0}(J)\,M^2(J+1,n)
+ K_{2,1}(J)\,M^2(J,n),  \nonumber \\
{\cal B} &=& \Delta(J)^{-1} \left[ 2 M^2(J+1,n) - M^2(J+2,n) -
M^2(J,n) \right], \nonumber \\
{\cal D} &=& M^2(J,n+1) - M^2(J,n), \nonumber \\
{\cal C} &=& K_{1,0}(J)\,\sqrt{J(J+1)}\,M^2(J+2,n) +
K_{2,1}(J)\,\sqrt{(J+1)(J+2)}\,M^2(J,n)\nonumber \\ 
&-&K_{2,0}(J)\,\sqrt{J(J+2)}\,M^2(J+1,n) -n \left[ M^2(J',n'+1) -
M^2(J',n')
\right],
\end{eqnarray}
where
\begin{equation}
\label{3.3}
\Delta(J) = 2\sqrt{J+1}-\sqrt{J+2}-\sqrt{J} \quad \text{and} \quad
K_{i,j}(J) = \frac{\sqrt{J+i}-\sqrt{J+j}}{\Delta(J)}.
\end{equation}
The interest of such formulae is that they are universal in the sense
that the quantities ${\cal A}$, ${\cal B}$, ${\cal C}$ and ${\cal D}$
are independent on $J$ and $n$. Given a flavor sector, they could be, in
principle, checked for various orbital and radial multiplets.
Moreover, the quantities ${\cal A}$ and ${\cal D}$ are even independent
on $m$, so that they can be checked independently in several flavor
sectors. Finally, the ratio ${\cal D}/{\cal A}$ depends on $f$ only and
thus can provide a strong test.

\par In principle, one could obtain physical quantities from the
experimental masses. Starting from ${\cal D}/{\cal A} = R(f)$, one can
determine the value of $f$ (see Fig.~\ref{fig:coef}). The quantities
$A(f)$, $B(f)$, $C(f)$, $D(f)$ and $E(f)$ can then be calculated. The
${\cal A}$ and ${\cal D}$ quantities can provide us with cross-checked
values of the string tension. If
$C_{\text{S}} = C_{\text{V}} = 0$, the ${\cal B}$ term gives the quark
mass $m$; if $C_{\text{S}} \neq 0$ and $C_{\text{V}} \neq 0$, only the
quark mass difference in two flavor sectors can be obtained. Calculating
parameter $\kappa$ is more involved.

\par Unfortunately, our model is only valid for high values of $J$, and,
for such quantum numbers, a few ground states are known and none radial
excitation \cite{pdg}. Moreover, uncertainties exist about meson masses.
Nevertheless, we can try to obtain an estimation of parameter $f$ by
considering the mesons of the $\rho$ and $\phi$ families (see
Fig.~\ref{fig:regge}). 

\par The meson masses used to perform calculations are center of gravity
of meson multiplets whose members are characterized by an internal total
spin quantum number equal to one. They are calculated, as well as the
corresponding uncertainties, with the procedure given in
Ref.\cite{brau98}. Using the only three available masses $J$ = 1, 2 and
3 for the $\rho$ family and taking into account the error on these data,
we can determine the range possible for the function ${\cal A}=
a\,A(f)$. Assuming usual values for the string tension $a$, we can
deduce the possible range for $f$. Unfortunately, we obtain no
constraint on this parameter since values from 0 to 1 are compatible
with data. The same calculation done for the $\phi$ family leads to the
conclusion that $f$ must be very close to 0, that is to say that the
confinement is only of vector-type. We have remarked that the results
are very sensitive to the value of the masses. If other states with
higher values of $J$ were known, our conclusion could be changed. 

\par It is also possible to calculate $f$ by considering a meson and its
radial excitations (${\cal D}=2a\,E(f)$). This procedure is more
questionable since only the first radial excitation of the $J=1$ meson
is known in the $\rho$ and $\phi$ families. Nevertheless, if the
calculation is performed for the two families, we find that all values
for $f$ from 0 to 1 are compatible with the data. 

\section{Concluding remarks}
\label{sec:conc}

We have shown that light-light and heavy-light mesons exhibit linear
orbital and radial trajectories in the dominantly orbital states (DOS)
model. Slopes of both type of trajectories depend only on the string
tension and on the vector-scalar mixture in the confinement potential.
From our work, it turns out that small finite quark masses do not alter
significantly the linearity, specially in the case of dominant
vector-type confining potential. As expected, the Coulomb-like potential
has no effect on trajectories but its influence on the meson masses has
been determined in the approximation of the DOS description. Lastly, we
have shown that the ratio of radial to orbital energies depends on the
scalar-vector confinement mixture only.

\par Formula~(\ref{2.19}) gives the parameter dependence of square mass
in the limit of small masses and small strength of the short range
potential, or in the limit of large angular momentum. Outside these
limits, we have remarked that this formula only differ slightly from
exact results in a large range of parameters. This allows to apply our
approximation for physical situations and to compare our calculations
with experiment.

\par From data available, it is difficult to determine the value of the
parameter $f$, that is to say the scalar-vector mixture in the
confinement. Strictly speaking, our results are only semiclassical ones,
and experimental radially excited meson masses are only known
for small value of total spin. Nevertheless, our calculations favor
a dominant vector-type confining interaction. 

\par In our study, we completely neglect the quark spin. Actually, this
is not an important drawback. Regge and vibrational trajectories concern
only mesons with an assumed internal total spin quantum number equal to
one \cite{pdg}. In usual models the spin-dependent part of the potential
is the hyperfine interaction stemming from the one-gluon exchange
interaction (and may be from the vector part of the confinement)
\cite{luch91}. In other models, this spin-dependent part stems from an
instanton induced interactions \cite{sema95,blas90}. In both cases, the
contribution of the spin to the meson masses is small or vanishing for
mesons with total spin equal to one. 

\par In Sec.~\ref{ssec:heavy}, we mentioned a method to study the
heavy-light mesons in the limit of an infinitely massive heavy quark.
Such a method is used in Ref.~\cite{olss97}. It could also be
interesting to develop the calculations in the more general case of two
finite different masses.

\acknowledgments

C. Semay would like to thank the F.N.R.S. for financial support, and F.
Brau would like to thank the I.I.S.N. for financial support. B.
Silvestre-Brac is grateful to Mons-Hainaut University for financial
support, and for good working conditions.


\clearpage

\end{document}